\documentclass[12pt, twoside, here]{article}
\usepackage{epsf}
\usepackage{times,colordvi,amsmath,epsfig,float,color,multicol}
\usepackage{graphics}
\usepackage{hhline}
\usepackage[large]{subfigure}
\usepackage[latin1]{inputenc}
\usepackage{listings}
\usepackage{rotating}

\newtheorem{theorem}{Theorem}

\title{Quantification of the South African Lockdown Regimes, for the \mbox{SARS-CoV-2} Pandemic, and the Levels of Immunity They Require to Work}

\author{S. J. Childs \\ \\ {\small\em Department of Mathematics and Applied Mathematics , University of Fort Hare,} \\ {\small\em Private Bag X1314, Alice, 5700, South Africa}}

\date{2 July, 2020}       

\begin{document}

\maketitle

\begin{abstract}
\noindent {\em This research quantifies the various South African lockdown
regimes, for the \mbox{SARS-CoV-2} pandemic, in terms of the basic reproduction
number, $r_0$. It further calculates the levels of immunity required for these
selfsame lockdown regimes to begin to work, then predicts their perceived
values, should infections have been underestimated by a factor of 10. The first,
level-5 lockdown was a valiant attempt to contain the highly infectious,
SARS-CoV-2 virus, based on a limited knowledge. Its basic reproduction number
($r_0 = 1.93$) never came anywhere close to the requirement of being less than
unity. Obviously, it could be anticipated that the same would apply for
subsequent, lower levels of lockdown. The basic reproduction number for the
level-3 lockdown was found to be $2.34$ and that of the level-4 lockdown,
$1.69$. The suggestion is therefore that the level-4 lockdown might have been
marginally `smarter' than the `harder', level-5 lockdown, although its basic
reproduction number may merely reflect an adjustment by the public to the new
normal, or the ever-present error associated with data sets, in general. The
pandemic's basic reproduction number was calculated to be $3.16$, in the Swedish
context. The lockdowns therefore served to ensure that the medical system was
not overwhelmed, bought it valuable time to prepare and provided useful data.
The lockdowns nonetheless failed significantly in meeting any objective to
curtail the pandemic.} 
\end{abstract}

Keywords: Pandemic; \mbox{SARS-CoV-2}; Covid-19; basic reproduction number; threshold; SIR equations; epidemic; South Africa; Sweden.

Declaration of interest: None.

\section{Introduction}

South Africa can be thought of as lucky, in that it obtained reasonably advanced
warning of the impending pandemic, however, it is challenged in the way of its
living conditions and basic hygiene. Many of its citizens live in extremely close proximity to one another and they are in constant physical contact. 

The official narrative of the South African, \mbox{SARS-CoV-2} epidemic is that
it commenced around the 1st of March 2020, following the return of a group of
seven, infected tourists, from a skiing holiday in Italy \cite{tourists}.
Multiple sources (e.g. \cite{nicd} and \cite{openingWeeks}) subsequently
reported that, by the 11th of March, the number of cases had risen to 13. By the
15th of March, there were already 61 reported cases (\cite{data},
\cite{openingWeeks} and \cite{data3}). By the 27th of March, a total of 1170
cases had been recorded and 1138 of them were still active (\cite{data},
\cite{openingWeeks} and \cite{data3}). A cursory inspection of these early data
suggests that the first infection, in actual fact, took place substantially
before March, or at very least, recoveries in early March were not properly reported. 

The level-5 lockdown commenced on the 26th of March and ended on the 1st of May.
Citizens were prohibited from leaving their residences for any other reason than
to purchase food, electricity, fuel, or other items deemed essential. The rules
only allowed foodstuffs and other essential goods to be sold. Limits on the
numbers of customers, in the same shop at the same time, were also set. Only
those employed in supermarkets, medical personnel, emergency plumbers and their
like were allowed to continue working. The sale of tobacco and liquor was also
prohibited. Further details of the level-5 lockdown regulations are available at
\cite{lockdownRegulations}. The opening weeks of the level-5 lockdown were
characterised by migration and a substantial lack of compliance. This is borne
out by, for example, a plethora of photographs showing traffic jams at Hertzog
Bridge, in Aliwal North, and videos showing kilometres-long convoys of
minibuses, at other Eastern Cape borders; all of which appeared on social media.
The flagrant disregard for the rules was not limited to migrant workers and
holiday makers. The elite of at least one established institution had guests
around, tasted and distributed homemade liquor among themselves, had workmen in
and sent their children to each others' houses. In many areas, the large crowds
that queued outside supermarkets suggested that people needed time to adjust to
the level-5 rules. People in many of these queues were in physical contact, or
very close to it. By the 1st of May, the day after the level-5 lockdown had
ended and the day the level-4 lockdown commenced, a total of 5951 cases had been
recorded and 3453 of those were still active (\cite{data}, \cite{openingWeeks}
and \cite{data3}).

The level-4 lockdown commenced on the 1st of May and ended on the 31st day of
that same month. The level-4 rules allowed for exercise in public places,
between the hours of 06h00 and 09h00, stipulated that people must wear masks, at
all times, in public and a travel ban was implemented on the 7th of May. Further
details of the level-4 lockdown regulations are available at
\cite{lockdownRegulations}. This lockdown was characterised by fewer
infringements, as the public appeared to become adjusted to the new normal. By
the 31st of May, at the end of the level-4 lockdown, a total of 32683 cases had
been recorded and 15191 of those were still active (\cite{data},
\cite{openingWeeks} and \cite{data3}).

The level-3 lockdown commenced on the 1st of June and was still in force on the
first day of that following month. The level-3 rules permitted the sale of
liquor and, on the 8th of June, grades 7 and 12 returned to school. On the night
of the 17th of June, so-called ``advanced level-3'' was announced. Hairdressers,
restaurants, accredited and licenced accommodation, cinemas and casinos were
notified that they could resume operating. How long they took to respond and to
what level customers resumed patronising these establishments, is not known.
Further details of the level-3 lockdown regulations are available at
\cite{lockdownRegulations}. By the 1st of July, a total of 159333 cases had been
recorded and 80559 of those were still active (\cite{data}, \cite{openingWeeks}
and \cite{data3}).

Sweden set no special regulations for their \mbox{SARS-CoV-2} epidemic and their
data (\cite{swedenData} and \cite{moreSwedenData}) therefore serve as a
convenient, nonetheless, very approximate control for South Africa's lockdown
experiments. Sweden's climate, their population's way of life, their
population-densities etc. are, of course, all very different to those in South
Africa, so one has to exercise caution in drawing any conclusions from the
comparison.

The basic reproduction number, $r_0$, may be defined as the total number of
people that an infected member of the population would manage to infect, before
recovering, in an otherwise naive population. Its significance is, however, far
greater. In the basic reproduction number one has a holistic quantity with which
to characterise both the infectiousness of a disease, as well as the environment
in which it propagates, right down to things like the temperature of nasal
passages, the rules of a lockdown and even the level of non-compliance. It
allows for the lockdown-threshold to be calculated, which is the minimum level
of immunity the population must have in order for the epidemic to downgrade to a
mere disease, as well as the level at which all infection ceases. In one Chinese
context, the basic reproduction number for the \mbox{SARS-CoV-2} virus was found to be 2.2 \cite{meanIncubationPeriod}.

In this research, the basic reproduction number, associated with each lockdown
regime, is calculated from the numbers of active infections and the total tallys
of all infections since the outbreak of the pandemic. Not only have so-called
individual level models, such as those of \cite{andersonAndMay} and
\cite{ilmVersusPlm}, been discredited by \cite{ilmVersusPlm} as a means for
calculating $r_0$, they are also far more laborious than the simple method used
in this work. The various lockdown-thresholds, as well as the points at which
all infection would cease, are, in turn, calculated from these basic
reproduction numbers. Finally, the perceived values of these selfsame quantities
are predicted, should infections have been underestimated by a factor of 10 and
these results are compelling. 

\section{Derivation of the Relevant Formulae}

Kermack and McKendrick's SIR equations \cite{kermackAndMcKendrick} state that
\begin{eqnarray} 
\frac{d S}{dt} &=& - \beta S(t) I(t) \label{kermackAndMcKendrick1st} \\ 
\frac{d I}{dt} &=& \beta S(t) I(t) - \gamma I(t) \label{kermackAndMcKendrick2nd} \\
\frac{d R}{dt} &=& \gamma I(t), \nonumber
\end{eqnarray}
in which $\beta$ denotes the rate of potentially infectious encounters to which
a member of the population is exposed, $S(t)$ denotes the susceptible fraction
of the population, $I(t)$ denotes the infected fraction of the population,
$\gamma$ denotes the combined rate of recovery and death, while $R(t)$ denotes
the `resistant' fraction of the population, those that have either acquired
immunity, or died from the disease. The characterisation of an epidemic in
terms of a basic reproduction number, $r_0$, the calculation of $r_0$, the
threshold and the point at which an infection completely burns itself out, are
ultimately all based on these equations.
\begin{theorem}[Basic Reproduction Number]\label{rNoughtTheorem} An epidemic is only possible for $r_0 > 1$.
\end{theorem}
{\bf Proof:} By definition, a disease is not an epidemic unless the level of infection increases. From Equation (\ref{kermackAndMcKendrick2nd}), one observes
\[
\frac{d I}{dt} > 0 \iff \beta S(t) - \gamma > 0 \iff \frac{\beta}{\gamma} S(t) > 1
\]
Since $S(t) \le 1$ for all $t$,
\[
\frac{d I}{dt} > 0 \Rightarrow r_0 > 1,  
\]
in which $r_0 = \frac{\beta}{\gamma}$, concluding the proof.

In $r_0$ one therefore has a holistic quantity with which to characterise both the infectiousness of a disease, as well as the environment in which it propagates, right down to things like the temperature of nasal passages, the rules of a lockdown and even the level of non-compliance. The so-called $r$-effective, $r_0 \times S(t)$, is a characterisation of the disease's infectiousness, pertinent to a given point in time, as the epidemic progresses, or where immunity is present.

\subsubsection*{A Formula With Which to Recover the Basic Reproduction Number  From the Data}

Using the chain rule,    
\begin{eqnarray*}
\frac{dI}{dS} &=& \frac{dI}{dt} \frac{dt}{dS}, 
\end{eqnarray*}
then substituting Equations (\ref{kermackAndMcKendrick2nd}) and (\ref{kermackAndMcKendrick1st}), the expression,
\begin{eqnarray*}
\frac{dI}{dS} &=& - 1 + \frac{\gamma}{\beta S} \\
&=& - 1 + \frac{1}{r_0 S(t)},
\end{eqnarray*}
is obtained. Integrating over S,
\begin{eqnarray} \label{infectives}
\int \frac{dI}{dS} \ dS &=& - S + \frac{1}{r_0} {\mathop {\rm ln}} S + c \nonumber \\
\Rightarrow I(t) &=& - S(t) + \frac{1}{r_0} {\mathop {\rm ln}} \left[ S(t)  \right] + c,
\end{eqnarray}
is obtained, in which $c$ is the unknown constant of integration. At some $t = t_i$, 
\[
c = I(t_i) + S(t_i) - \frac{1}{r_0} {\mathop {\rm ln}} \left[ S(t_i) \right].
\]
Substituting this back into Equation (\ref{infectives}),
\begin{eqnarray*}
I(t) &=& - S(t) + \frac{1}{r_0} {\mathop {\rm ln}} \left[ \frac{S(t)}{S(t_i)} \right] + I(t_i) + S(t_i), \\
\end{eqnarray*}
is obtained. Evaluating this equation over some time interval $[ t_1, t_2 ]$ and solving for $r_0$, 
\begin{eqnarray} \label{formula}
r_0 = \frac{ {\mathop {\rm ln}} \left[ \frac{S(t_2)}{S(t_1)} \right] }{ \left[ I(t_2) + S(t_2) - I(t_1) - S(t_1) \right] }.
\end{eqnarray}
Notice that this formula is robust against any movement of $I(t)$ by an additive
constant, up or down. Such movement has no effect on the calculation of $r_0$,
whatsoever. In fact, the formula, Equation (\ref{formula}), is reasonably robust
against any data error that does not effect the relative values, or slopes of
the functions concerned. 

Once $r_0$ has been calculated, $S(t_2) = S_\infty$, can be recovered from this
selfsame equation, by considering that $S_\infty$ is the point at which all
infection ceases, i.e. by setting $I(t_2)=0$, in the above equation. Once the level of infectiousness for a given lockdown has been characterised in
terms of the basic reproduction number, $r_0$, it is instructive to know the
level of immunity at which the lockdown renders the disease no longer an
epidemic. The levels to which susceptibility must drop, in order for the
relevant lockdown regimes to begin to work, can be determined by the application
of the threshold theorem.
\begin{theorem}[The Threshold Theorem]\label{thresholdTheorem}
No epidemic occurs when $S(t) < \frac{1}{r_0}$.
\end{theorem}
{\bf Proof:} By definition, a disease is not an epidemic if the level of infection decreases. From Equation (\ref{kermackAndMcKendrick2nd}), one observes
\[
\frac{d I}{dt} < 0 \iff \beta S(t) - \gamma < 0 \iff
S(t) < \frac{\gamma}{\beta}.
\]
Since the quantity on the right is immediately recognisable as $\frac{1}{r_0}$, the above statement may be more concisely expressed as,
\[
\frac{d I}{dt} < 0 \iff S(t) < \frac{1}{r_0},
\]
concluding the proof. 

In other words, if the susceptible fraction of the population is still above
$\frac{1}{r_0}$, for a given lockdown, that lockdown will not serve to curtail
the epidemic, only to delay it. In such circumstances, the epidemic will only temporarily slow during the lockdown, then resume, as before, after it. Only at the threshold does $r$-effective drop to unity.

\section{Asymptomatic, or Undiagnosed \mbox{SARS-CoV-2} Infections}

What of the large number of asymptomatic, or undiagnosed \mbox{SARS-CoV-2}
infections, not reflected in the data? The head of the CDC, Robert Redfield's
opinion on the topic of asymptomatic or undiagnosed \mbox{SARS-CoV-2}
infections, in the U.S.A., is that antibody testing reveals that ``A good rough
estimate now is 10 to 1'' \cite{redfield} and others, in similar positions all over the world, have expressed similar sentiments. 

Suppose one were to determine that both $I(t)$ and therefore, the resistant
portion arising from the current epidemic, $R(t) - R(0)$, are in actual fact
higher by a factor, $a$. That is, 
\[
I(t) = a {\tilde I}(t) \hspace{10mm} \mbox{and} \hspace{10mm} R(t) = a \left[ {\tilde
R}(t) - R(0) \right] + R(0), 
\]
in which the tilde denotes the incorrectly measured, data-value in an epidemic
that begins at \mbox{$t=0$}. Then
\begin{eqnarray} \label{realS} 
S(t) &=& 1 - a \left[ {\tilde I}(t) + \left( {\tilde R}(t) - R(0) \right)  \right] - R(0) \nonumber \\
&=& 1 - a + a \left[ 1 - \left( {\tilde I}(t) + {\tilde R}(t) \right) \right] - ( 1 - a ) R(0) \nonumber \\
&=& a {\tilde S}(t) + ( 1 - a ) [ 1 - R(0) ], 
\end{eqnarray}
in which ${\tilde S}(t)$ denotes the perceived, incorrect-data-based value of
$S(t)$. The threshold therefore occurs at  
\[
a {\tilde S}(t) + ( 1 - a ) [ 1 - R(0) ] = 1 / r_0,  
\]
according to Theorem \ref{thresholdTheorem}, from which the corresponding, perceived, incorrect-data-based threshold can be recovered by changing the subject of the above equation to 
\[
{\tilde S}(t) = \frac{1}{a} \left[ \frac{1}{r_0} + ( a - 1 ) \left[ 1 - R(0) \right] \right]. 
\]
Changing the subject of Equation (\ref{realS}) also facilitates the calculation of the perceived, incorrect-data-equivalent of $S_\infty$, 
\begin{eqnarray*}
{\tilde S}_\infty &=& \frac{ S_\infty + ( a - 1 ) \left[ 1 - R(0) \right] }{a}.
\end{eqnarray*}
Of course, $R(0) = 0$ for a novel infection such as \mbox{SARS-CoV-2}; or so one believes.
 

\section{Fitting Curves to the Data}

Curves are fitted to a level-5 data set, a level-4 data set, a level-3 data set
and a Swedish data set. Epidemiological data are usually presented in the format
``numbers of current infections'' and ``total number of cases''. The present
case of the \mbox{SARS-CoV-2} pandemic is no exception. If $N$ is the size of
the population, the aforementioned quantities are just $I(t) N$ and
\mbox{$[I(t)+R(t)] N$}, respectively; always assuming that the population is
naive in the case of a novel infection like \mbox{SARS-CoV-2} (i.e. that $R(0) =
0$). This standard, epidemiological data format merely implies an additional
step; namely that the values S(t) and I(t), to be used in the formulation
(\ref{formula}), must first be calculated from 
\begin{eqnarray} \label{dataConversion}
S(t) = \frac{ N - [I(t) + R(t)] N } {N} \hspace{10mm} \mbox{and} \hspace{10mm} I(t) = \frac{I(t) N}{N}.
\end{eqnarray}
In 2020, population sizes were estimated to be \mbox{59 140 502} and
\mbox{10 089 108} for South Africa \cite{saPopulation} and Sweden
\cite{swedishPopulation}, respectively. 

\subsubsection*{The Applicable Temporal Interval}

The mean incubation period for the \mbox{SARS-CoV-2} virus is 5.2 days with the
95th percentile occurring at 12.5 days \cite{meanIncubationPeriod}. The World
Health Organisation (WHO) quotes the incubation period as being anywhere between
2 and 10 days \cite{incubationPeriodWHO}, China's National Health Commission
(NHC) found symptoms to appear anywhere from 10 to 14 days after infection
\cite{incubationPeriodNHC}, the United States' Centres for Disease Control and
Prevention (CDC) found symptoms to appear anywhere from 2 to 14 days after
infection \cite{incubationPeriodCDC} and the Chinese, online DXY.cn quotes the
incubation period as being anywhere between 3 to 7 days after infection;
possibly as high as 14 days \cite{incubationPeriod}. Two, record outliers for
the incubation period are 19 days \cite{19days} and 27 days \cite{27days}. There
is therefore considerble agreement on a lower bound of no less than 2 days and
an upper bound of no more than 14 days for the incubation period. Allowing a
further day for diagnosis, data were used from the sixteenthth day after the
relevant lockdown began until the first day after it ended.

\subsection{The Level-5 Lockdown}

The level-5 lockdown commenced on the 26th of March and ended on the 1st of May. A period of 15 days was allowed for the viral incubation period and the subsequent diagnosis of an infection. It was also assumed that the termination of the level-5 lockdown would not reflect in the data for at least 24 hours. 
\begin{figure}[H]
    \begin{center}
        \includegraphics[width=15cm, angle=0, clip = true]{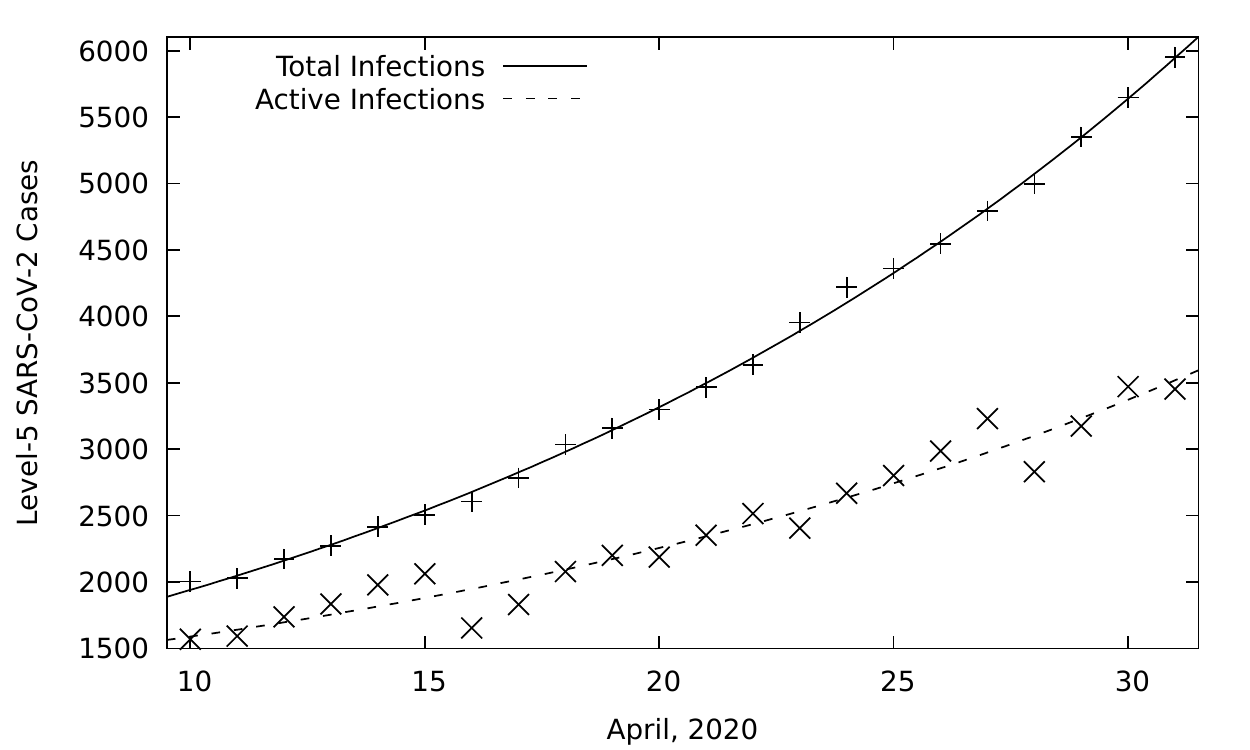}
	\caption{Level-5, best fits to \mbox{SARS-CoV-2}, infection data (10th of March to the 1st of May, 2020).} \label{level5}
   \end{center}
\end{figure}
Curves were accordingly fitted to the subset of data (\cite{data}, \cite{openingWeeks} and \cite{data3}) which commenced on the 10th of April and terminated on the 1st of May. The curves fitted to the data, using Gnuplot, are depicted in Figure \ref{level5}. The formula for the ``total infections'' curve is 
\[ 
1182.86 \ e^{0.0523907 \ t} - \ 57.0624
\] 
and the formula for the ``active infections'' curve is 
\[ 
594.027 \ e^{0.0513925 \ t} + \ 595.899. 
\] 
The values these formulae yield are provided in \mbox{Table \ref{results}}. They are first substituted into Equations (\ref{dataConversion}), which, in turn, provide the necessary inputs for Equation (\ref{formula}).  

\subsection{The Level-4 Lockdown}

\begin{figure}[H]
    \begin{center}
        \includegraphics[width=15cm, angle=0, clip = true]{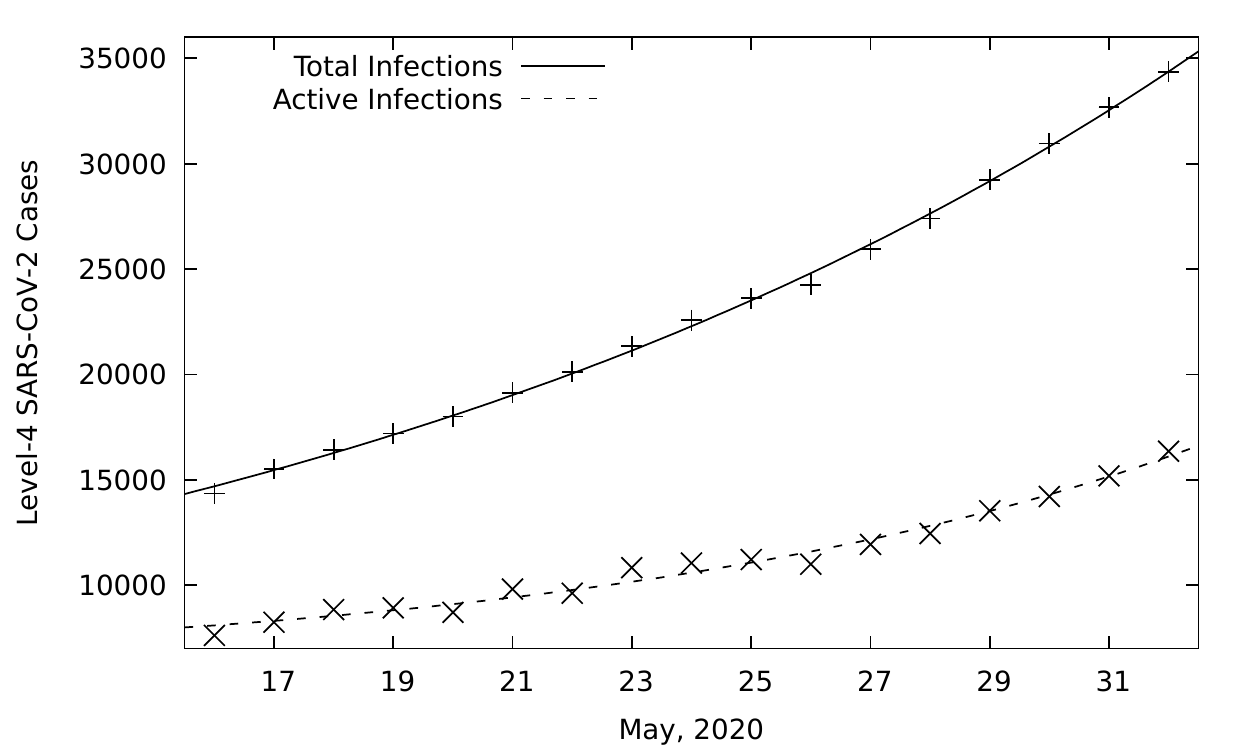}
	\caption{Level-4, best fits to \mbox{SARS-CoV-2}, infection data (16th of May to the 1st of June, 2020).} \label{level4}
   \end{center}
\end{figure}

The level-4 lockdown commenced on the 1st of May and ended on the 31st of May. Once again, a period of 15 days was allowed for the viral incubation period and the subsequent diagnosis of an infection. Once again, it was also assumed that the termination of the level-4 lockdown would not reflect in the data for at least 24 hours. Curves were accordingly fitted to the subset of data (\cite{data}, \cite{openingWeeks} and \cite{data3}) which commenced on the 16 of May and terminated on the 1st of June. The curves fitted to the data, using Gnuplot, are depicted in Figure \ref{level4}.
The formula for the ``total infections'' curve is 
\[ 
5073.89 \ e^{0.0580619 \ t} + \ 1848.01
\] 
and the formula for the ``active infections'' curve is 
\[ 
426.93 \ e^{0.0988331 \ t} + \ 6020.24. 
\] 
The values these formulae yield are provided in \mbox{Table \ref{results}}. They are first substituted into Equations (\ref{dataConversion}), which, in turn, provide the necessary inputs for Equation (\ref{formula}). 

\subsection{The Level-3 Lockdown}

\begin{figure}[H]
    \begin{center}
        \includegraphics[width=15cm, angle=0, clip = true]{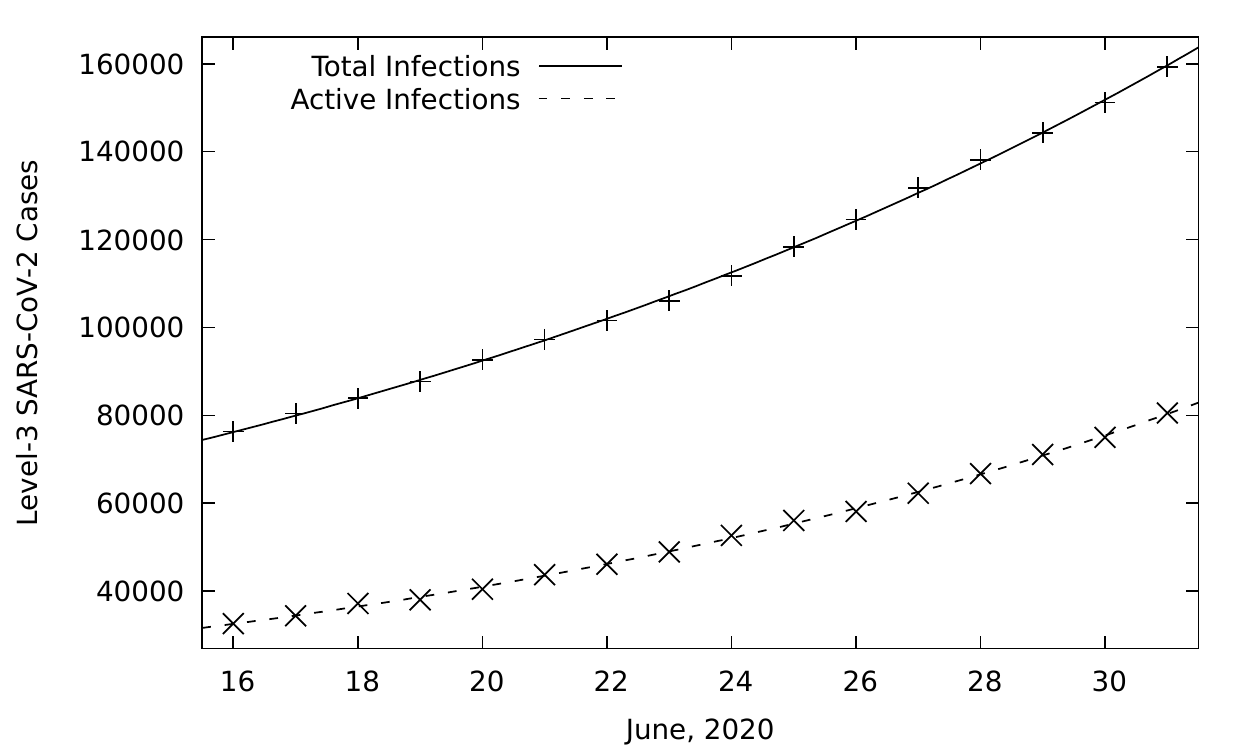}
	\caption{Level-3, best fits to \mbox{SARS-CoV-2}, infection data (16th of June to the 1st of July, 2020).} \label{level3}
   \end{center}
\end{figure}

The level-3 lockdown commenced on the 1st of June and was still in force one month later. Once again, a period of 15 days was allowed for the viral incubation period and the subsequent diagnosis of an infection. 
Curves were accordingly fitted to the subset of data (\cite{data}, \cite{openingWeeks} and \cite{data3}) which commenced on the 16 of June and terminated on the 1st of July. The curves fitted to the data, using Gnuplot, are depicted in Figure \ref{level3}.
The formula for the ``total infections'' curve is 
\[ 
29907 \ e^{0.0526168 \ t} + \ 6804.39
\] 
and the formula for the ``active infections'' curve is 
\[ 
9372.92 \ e^{0.067175 \ t} + \ 5123. 
\] 
The values these formulae yield are provided in \mbox{Table \ref{results}}. They are first substituted into Equations (\ref{dataConversion}), which, in turn, provide the necessary inputs for Equation (\ref{formula}). 

\subsection{Sweden}

Sweden's data \cite{swedenData} appear to be regularly subjected to major
revisions. The data used in this work might therefore no longer be current at
the time of reading, however, the actual values of the data and consequentlly
the results, change little. The regular revisions suggest that someone cares. 
\begin{figure}[H]
    \begin{center}
        \includegraphics[width=15cm, angle=0, clip = true]{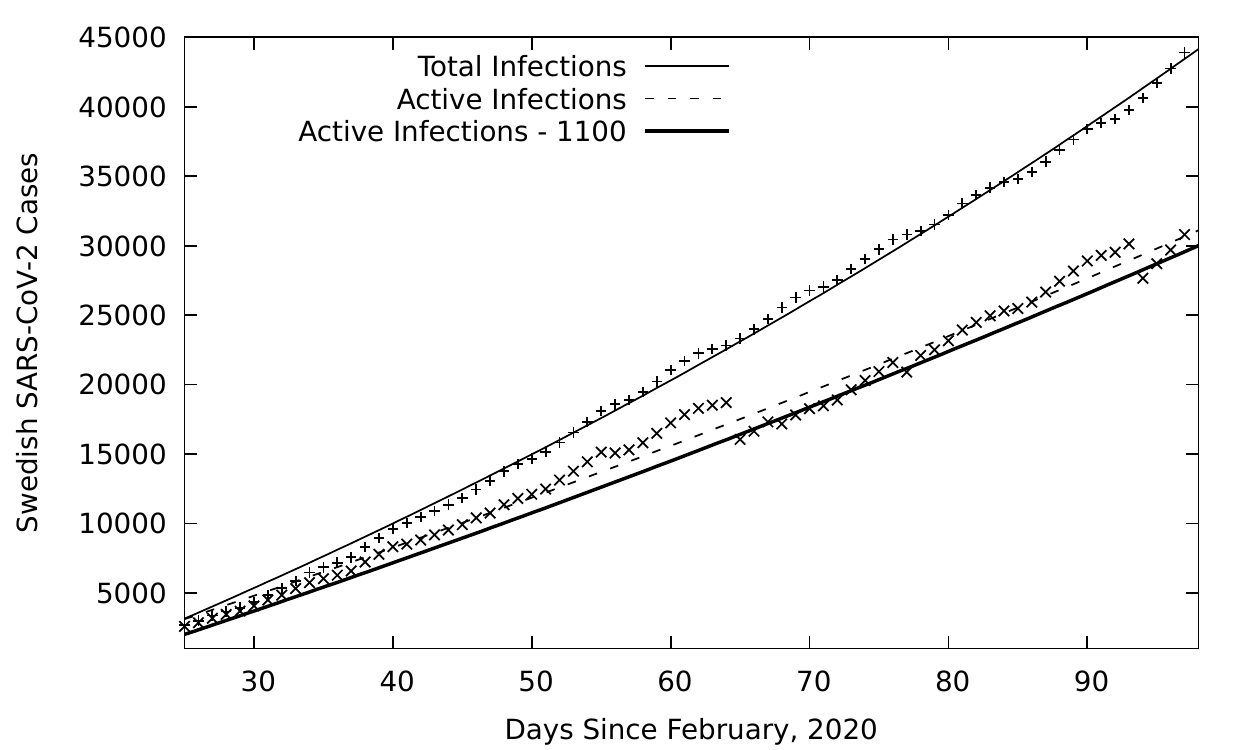}
	\caption{Best fits to the Swedish, \mbox{SARS-CoV-2}, infection data, from the 15th of March to the 5th of June, 2020.} \label{swedenAll}
   \end{center}
\end{figure}
Visual inspection of the Swedish data revealed a much less prominent, initial
step, up. It is fairly safe to assume this would be indicative of discovery,
rather than growth of the actual \mbox{SARS-CoV-2} pandemic, itself, as they had
no lockdowns. Sweden's data (\cite{swedenData}) are slightly problematic, in
that an uncharacteristically large number of recoveries were reported on the 4th
of May and the 2nd of June. The synchronicity in these recoveries is highly
suspect and they manifest themselves as two, visibly large steps, down, in the
graph of ``active infections'' (Figure \ref{swedenAll}). One suspects that these
`saw-teeth' are, in actual fact, artefacts, which arose due to a delay in the
diagnosis and reporting of recoveries, a suspicion which seemed to originally be
corroborated by \cite{swedenData}. It is strongly suggestive of an incorrect
record of ``active infections'' prior to the day in question. A decision was
therefore made to fit curves to as much data as possible, barring early March
(for fear that it documented discovery, rather than growth). Of course, one has
to wonder whether the data subsequent to the 3rd of May and the 2nd of June were
not also afflicted by a similar hoarding of recoveries and, indeed, they seem to
be \cite{swedenData}. The `saw-teeth' had the effect of `lifting' the curve
slightly, however, the overall slope appeared to be and, indeed, should mostly
be correct for the temporal period selected. Fortunately, the formula for $r_0$,
\mbox{Equation (\ref{formula})}, is robust against any movement of $I(t)$ by an
additive constant, up or down. The $r_0$ calculated according to the formula,
\mbox{Equation (\ref{formula})}, does not change in any way, whatsoever, for
such movement of $I(t)$. A corrected curve, obtained by revising the original
fit downward by 1100 infections, is nonetheless provided for the reader's
benefit, in Figure \ref{swedenAll}.   

The curves fitted to the subset of data (\cite{swedenData} and
\cite{moreSwedenData}), which commenced on the 15th of March and terminated on
the 5th of June, are depicted in \mbox{Figure \ref{swedenAll}}. The curves were
fitted using Gnuplot. The formula for the ``total infections'' curve is 
\[ 
54205 \ e^{0.0067666 \ t} - \ 61051.4
\] 
and the formula for the ``active infections'' curve is 
\[ 
82641.9 \ e^{0.00368722 \ t} - \ 87513.5.
\] 
The values these formulae yield are provided in \mbox{Table \ref{results}}. They are first substituted into Equations (\ref{dataConversion}), which, in turn, provide the necessary inputs for Equation (\ref{formula}). 


\section{Results}

The results, as well as the inputs from which they were obtained, are provided
in \mbox{\mbox{Table \ref{results}}} and \mbox{\mbox{Table \ref{results2}}}, on \mbox{page \pageref{results}}. 


\begin{sidewaystable}
\begin{tabular}{|c||c|c|c||c|c|c||c||c||c||} \hline & & & & & & & & & \\ Regime & Date 1 & Total Cases & Active Cases & Date 2 & Total Cases & Active Cases & $r_{0}$ & Threshold $S \ / \%$ & $S_\infty \ / \%$ \\ & & & & & & & & & \\ \hline
level 5 & 10.04 & 1 940 & 1 589 & 01.05 & 5 945 & 3 518 & 1.93 & 51.8 & 22.3 \\ \hline 
level 4 & 16.05 & 14 695 \ & 8 096 & 01.06 & 34 376 \ & 16 110 \ & 1.69 & 59.3 & 31.4 \\ \hline  
level 3 & 16.06 & 76210 \ & 32580 & 01.07 & 159617 \ & 80330 \ & 2.34 & 42.7 & 13.0 \\ \hline 
Sweden & 26.03 & 3580 & 3443 & 05.06 & 43443 \ & 30663 \ & 3.16 & 31.6 & \ 5.0 \\ \hline 
\end{tabular}
\caption{The basic reproduction number, $r_0$, the consequent threshold and the
point at which all infection ceases, $S_\infty$. With the exception of Sweden,
all are calculated from data associated with the relevant lockdown regimes,
imposed for the \mbox{SARS-CoV-2} pandemic, in South Africa.} \label{results}
\vspace{20mm}
\begin{tabular}{|c||c|c|c||c|c|c||c||c||c||} 
\hline 
& & & & & & & & & \\ 
Regime & Date 1 & Total Cases$^\dagger$\footnotetext[0]{$\dagger$ The perceived, incorrect data.} & Active Cases$^\dagger$ & Date 2 & Total Cases$^\dagger$ & Active Cases$^\dagger$ & $r_{0}$ & Threshold ${\tilde S} \ / \%$ & ${\tilde S}_\infty \ / \%$ \\ 
& & & & & & & & & \\ 
\hline
level 5 & 10.04 & 1 940 & 1 589 & 01.05 & 5 945 & 3 518 & 1.93 & 95.2 & 92.2 \\ \hline 
level 4 & 16.05 & 14 695 \ & 8 096 & 01.06 & 34 376 \ & 16 110 \ & 1.69 & 95.9 & 93.1 \\ \hline  
level 3 & 16.06 & 76 210 \ & 32 580 & 01.07 & 159 617 \ & 80 330 \ & 2.39 & 94.2 & 91.2 \\ \hline 
Sweden & 26.03 & 3580 & 3443 & 05.06 & 43443 \ & 30663 \ & 3.23 & 93.1 & \ 90.5 \\ \hline 
\end{tabular}
\caption{The basic reproduction number, $r_0$ for $a = 10$, the consequent
perceived threshold and the perceived point at which all infection ceases. With
the exception of Sweden, all are calculated from data associated with the
relevant lockdown regimes, imposed for the \mbox{SARS-CoV-2} pandemic, in South
Africa.} \label{results2}
\end{sidewaystable}

\section{Conclusions}

There is very little difference between perceived and actual basic reproduction
numbers, should infections have been underestimated by a factor as high as 10.
They rise only very slightly. There is, however, a profound difference between
the actual and perceived, data-based thresholds and the susceptibility levels at
which the infection will vanish (comparing Tables \ref{results} and
\ref{results2}). The results  based on the assertion that infections have been
underestimated by a factor of 10, are compelling.

Although the initial attempt to contain the epidemic failed, the sequence of
lockdowns provided vital data to determine their associated basic reproduction numbers and, consequently, at what levels of immunity these selfsame lockdowns would begin to become effective in the future (their associated epidemic-thresholds). 

Early data, collected prior to the level-5 lockdown, probably document the
trajectory of discovery, rather than they do growth of the actual
\mbox{SARS-CoV-2} pandemic, itself. This conclusion is based on a discernible
step, up, in the graph, coupled with an observation that such phenomenal growth
would have implied an unrealistic basic reproduction number; one of somewhere
around 36! At very least, recoveries in the early data were under-reported. A
lack of confidence in the very early data, along with abrupt transitions in
lockdown regimes and their consequent inflection points (delayed, diffuse or
otherwise), necessitated that curves only be fitted locally. 

The first, level-5 lockdown was a valiant attempt to contain the highly
infectious, \mbox{SARS-CoV-2} virus, based on a limited knowledge. Its basic
reproduction number ($r_0 = 1.93$) never came anywhere close to the requirement
of being less than unity. To put this in context, most influenza epidemics have
a substantially lower $r_0$ than this virus has under the conditions of a
level-5 lockdown. Such a level-5 lockdown would only become efficacious in
curtailing the \mbox{SARS-CoV-2} pandemic, were it to be implemented around the
51.8 \% susceptibility level (\mbox{Table \ref{results}}). In other words, only
after 48.2 \% of the population has been infected. Under level-5 lockdown
conditions, the \mbox{SARS-CoV-2} virus would only vanish as a disease, forever,
at around the 22.3 \% susceptibility level (\mbox{Table \ref{results}}). If,
however, the assertion that infections have been underestimated by a factor of
10 is correct, then the susceptibility level for the above threshold will be
perceived as 95.2 \%, while the susceptibility level at which the infection will
vanish, forever, will be perceived as 92.2 \% (\mbox{\mbox{Table \ref{results2}}}).

Obviously, it could be anticipated that the basic reproduction number for
subsequent, lower levels of lockdown would also not come anywhere near the
requirement of being less than unity. Indeed, the basic reproduction number for
the level-3 lockdown was found to be $2.34$ and that of the level-4 lockdown,
$1.69$ (\mbox{Table \ref{results}}). It is nonetheless surprising that the value
characterising level-4 is not higher than that chracterising level-5. It could
reflect an adjustment by the public to the `new normal', the wearing of masks,
the travel ban, or any combination of these. The suggestion is therefore that
the level-4 lockdown was `smarter' than the 'harder' level-5 lockdown, although
the data may not be of sufficient quality to draw such a conclusion. 

The level-4 lockdown would only be efficacious in curtailing the pandemic, were
it to be implemented around the 59.3 \% susceptibility level (Table
\ref{results}). In other words, only after 40.7 \% of the population has been
infected. Under level-4 lockdown conditions, the \mbox{SARS-CoV-2} virus would
only vanish as a disease, forever, at around the 31.4 \% susceptibility level
(\mbox{Table \ref{results}}). If, however, the assertion that infections have
been underestimated by a factor of 10 is correct, then the susceptibility level
for the above threshold will be perceived as 95.9 \%, while the susceptibility
level at which the infection will vanish, forever, will be perceived as 93.1 \%
(\mbox{Table \ref{results2}}).

The level-3 lockdown would only be efficacious in curtailing the pandemic, were
it to be implemented around the 42.7 \% susceptibility level (Table
\ref{results}). In other words, only after 57.3 \% of the population has been
infected. Under level-3 lockdown conditions, the \mbox{SARS-CoV-2} virus would
only vanish as a disease, forever, at around the 13.0 \% susceptibility level
(\mbox{Table \ref{results}}). If, however, the assertion that infections have
been underestimated by a factor of 10 is correct, then the susceptibility level
for the above threshold will be perceived as 94.2 \%, while the susceptibility
level at which the infection will vanish, forever, will be perceived as 91.0 \%
(\mbox{Table \ref{results2}}).

In the Swedish context, the basic reproduction number for the \mbox{SARS-CoV-2}
pandemic was difficult to determine due to two `saw-teeth' in the data and the
author was tempted to use only the troughs, nonetheless, didn't. The basic
reproduction number was calculated to be 3.16 (\mbox{Table \ref{results}}) and
recent data suggests it has shot up, if it can be relied on. One also has to
exercise caution in drawing conclusions from any comparison with South Africa.
Sweden's population-densities, their climate, their population's way of life,
etc. are very different to those in South Africa. One can nonetheless conclude
that the $r_0$ for the \mbox{SARS-CoV-2} virus is high.

As much as a large proportion of the South African population live cheek by
jowl, it is no credit to them that all lockdowns were characterised by a
substantial lack of compliance. When one considers that similar lockdown regimes
in countries like New Zealand and, initially, Australia produced the desired
results, South Africans, to a certain extent need to blame themselves, as much
as they do their circumstances, living conditions and government. The lockdowns
served to ensure that the medical system was not overwhelmed, bought it valuable
time to prepare and provided useful data. It should also be remembered that a
pathogen essentially commits suicide by killing its host and temporarily
suppressing $r_0$ can facilitate genetic drift in one that mutates rapidly
enough. The lockdowns nonetheless failed significantly in meeting any objective
to curtail the \mbox{SARS-CoV-2} pandemic.

Despondence in the case of the \mbox{SARS-CoV-2} virus can be tempered, to a
limited extent, by contemplating the `flattening of the curve' in the context of
the Small-Epidemic Theorem. Severely reducing the infected fraction of the
population just above the threshold could possibly alter the outcome. Knowing the exact, incorrect-data value, or perceived position, of this threshold could therefore be considered reasonably important and some statistician urgently needs to do some random testing to determine what the exact value of the multiplicative factor, $a$, really is. 

\bibliography{covidLockdownBibliography}

\begin{thebibliography}{10}

\bibitem{incubationPeriodNHC}
China's national health commission news conference on coronavirus.
\newblock Al Jazeera.
\newblock \mbox{26 January, 2020}.

\bibitem{27days}
Coronavirus incubation could be as long as 27 days, {C}hinese provincial
  government says.
\newblock Reuters.
\newblock \mbox{22 February, 2020}.

\bibitem{data3}
Covid 19 za dashboard.
\newblock
  http://datastudio.google.com/u/0/reporting/1b60bdc7-bec7-44c9-ba29-be0e043d8534/page/hrUIB.
\newblock June, 2020.

\bibitem{redfield}
The mercury news.
\newblock
  http://www.mercurynews.com/2020/06/26/millions-may-have-had-coronavirus-in-the-past-witjout-knowing-it-cdc-says/.
\newblock \mbox{26 June, 2020}.

\bibitem{nicd}
National institute for communicable diseases.
\newblock
  http://www.nicd.ac.za/latest-confirmed-cases-of-covid-19-in-south-africa/.
\newblock 11 March, 2020.

\bibitem{tourists}
Sa people news.
\newblock
  http://www.sapeople.com/2020/03/09/coronavirus-cases-in-south-africa-rise-to-7-amongst-friends-from-italy-ski-holiday/.
\newblock 9 March, 2020.

\bibitem{lockdownRegulations}
South african government.
\newblock http://www.gov.za/coronavirus/guidelines.
\newblock May, 2020.

\bibitem{openingWeeks}
Wikipedia.
\newblock http://en.m.wikipedia.org/wiki/COVID-19\_pandemic\_in\_South-Africa.
\newblock \mbox{June, 2020}.

\bibitem{moreSwedenData}
Wikipedia.
\newblock http:/en.m.wikipedia.org/wiki/COVID-19\_pandemic\_in\_Sweden.
\newblock \mbox{June, 2020}.

\bibitem{data}
Worldometer.
\newblock http://www.worldometers.info/coronavirus/country/south-africa.
\newblock \mbox{June, 2020}.

\bibitem{swedenData}
Worldometer.
\newblock http://www.worldometers.info/coronavirus/country/sweden.
\newblock \mbox{June, 2020}.

\bibitem{saPopulation}
Worldometer.
\newblock
  http://www.worldometers.info/world-population/south-africa-population.
\newblock \mbox{April, 2020}.

\bibitem{swedishPopulation}
Worldometer.
\newblock http://www.worldometers.info/world-population/sweden-population.
\newblock \mbox{April, 2020}.

\bibitem{incubationPeriod}
Worldometer.
\newblock
  http://www.worldometers.info/coronavirus/coronavirus\-incubation\-period/.
\newblock \mbox{May, 2020}.

\bibitem{incubationPeriodWHO}
Novel coronavirus (2019-ncov) situation report-7.
\newblock Technical report, World Health Organisation, 2020.

\bibitem{incubationPeriodCDC}
Symptoms of coronavirus.
\newblock Technical report, Centres for Disease Control and Prevention, 2020.

\bibitem{andersonAndMay}
R.~Anderson and R.~May.
\newblock {\em Infectious Diseases of Humans}.
\newblock {O}xford University Press., 1992.

\bibitem{19days}
Yan Bai, Lingsheng Yao, and Tao~Wei et~al.
\newblock Presumed asymptomatic carrier transmission of covid-19.
\newblock {\em JAMA}, 323(14):1406--1407, 2020.

\bibitem{ilmVersusPlm}
Romulus Breban, Raffaele Vardavas, and Sally Blower.
\newblock Theory versus data: How to calculate $r_0$.
\newblock {\em PLoS one}, 2(3) e282, 2007.

\bibitem{kermackAndMcKendrick}
W.~O. Kermack and A.~G. McKendrick.
\newblock A contribution to the mathematical theory of epidemics.
\newblock {\em Proceedings of the Royal Society A}, 115:700--721, 1927.

\bibitem{meanIncubationPeriod}
Qun Li, Xuhua Guan, Peng Wu, Xiaoye Wang, Lei Zhou, Yeqing Tong, Ruiqi Ren,
  Kathy Leung, Eric Lau, Jessica Wong, Xuesen Xing, and Nijuan~Xiang et~al.
\newblock Early transmission dynamics in wuhan, china, of novel
  coronavirus-infected pneumonia.
\newblock {\em The New England Journal of Medicine}, 382:1199--1207, 2020.

\end{thebibliography}

\end{document}